\documentstyle[twoside,fleqn,npb,epsfig]{article}
\newcommand{\be}{\begin{equation}}
\newcommand{\ee}{\end{equation}}
\newcommand{\ba}{\begin{eqnarray}}
\newcommand{\ea}{\end{eqnarray}}
\newcommand{\baz}{\begin{eqnarray*}}
\newcommand{\eaz}{\end{eqnarray*}}

\title{Nucleon  Mass Corrections to Spin Dependent Structure Functions 
and  Relations Between their Twist-3 Contributions}

\author{Johannes Bl\"umlein\address{
 DESY Zeuthen, D-15738 Zeuthen, Germany} and 
Avto Tkabladze$^{\rm a}$\thanks{Alexander
von Humboldt Fellow}}

\begin{document}

\begin{abstract}
The nucleon mass corrections are calculated to all polarized 
structure functions for neutral and charged current deep inelastic
scattering in lowest order in the coupling constant. 
The impact of the target mass corrections on the general relations between
the twist--2 and twist--3 parts of the structure functions is studied and
three new relations between the twist--3 contributions are derived.
The size of nucleon mass corrections for the  $g_1$ and $g_2$ structure 
functions are estimated. 
\end{abstract}

\maketitle

\section{INTRODUCTION}

\noindent
In the  experiments in which polarized lepton scattering off 
polarized targets  has been studied so far the most data are taken  in 
the range of lower values of $Q^2$. In this domain nucleon--mass corrections
have to  be taken into account to analyze  the $Q^2$ dependence of structure
functions (SF).
The target mass effects of $O((M^2/Q^2)^k)$ form one contribution
to the power corrections. 
Unlike the case for dynamical higher twist 
effects the target mass corrections can be calculated in closed
form in all orders in $M^2/Q^2$ for deep inelastic SF. 

 Little is known so far on the relative strength of
dynamical higher twist operators and their scaling violations. Before
one may try to pursue a common treatment various conceptional problems
concerning higher twist operators have first to be solved. Due to this we 
limit the present analysis to a systematic study of the target mass 
corrections for all deep inelastic structure functions
extending earlier investigations~\cite{GP,PR,WA,MU}. 
From an experimental point of view, on the other hand,  the kinematic
higher twist effects emerge as a  background for the extraction 
of the dynamical higher twist terms at moderate energies.

Another goal of our investigation is to study the effect of the target 
mass corrections on the integral relations between  different
polarized structure functions in lowest order in the coupling constant.
 As in Ref.~\cite{BK} nucleon mass effects were 
disregarded, except of those implied by the kinematics in the Born cross
sections,  twist--3 contributions to the structure functions
$g_1, g_4$ and $g_5$ were not yet obtained. This picture has to be regarded as
partly incomplete,  since nucleon mass effects have either to be accounted
for thoroughly or to be neglected at all. In the latter case, however, 
the scattering cross sections for longitudinally polarized nucleons would 
not even contain the structure functions $g_2$ and $g_3$.

Two methods were proposed in the literature for the evaluation of the
target mass corrections. Nachtmann~\cite{Nachtmann} translated the usual
power--series expansion into an expansion of operators of definite spin.
The problem may be solved by applying the representation--theory of the
Lorentz group. A second method, which we used in our calculations, 
was proposed by  Georgi and Politzer~\cite{GP} 
and relies on resummation of the individual mass terms.

\section{RESUMMED EXPRESSIONS FOR THE STRUCTURE FUNCTIONS}

We consider the light-cone expansion of the forward Compton scattering 
amplitude (FCSA) in
momentum space. As a result, the $T$-product of two charged currents can be expressed through one quarkonic  operator at  leading order QCD,
$\Theta^{\pm\beta\{\mu_1,\cdots\mu_n\}}=
\bar q\gamma_5\gamma_{\beta}D_{\mu_1}\cdots D_{\mu_n}q$. The sign $\pm$ in the
 $\Theta$-operators corresponds to the charged currents combinations
\ba
\label{eqCOM}
T_{\mu\nu}^{\pm}(q^2, \nu) = T_{\mu\nu}^{W^-}(q^2, \nu)
                        \pm T_{\mu\nu}^{W^+}(q^2, \nu)~.
\ea
For neutral currents all expressions are the same as for  $T^{+}_{\mu\nu}$.
Decomposing the $\Theta^{\pm}$ operators into a symmetric part and 
a remainder 
one can isolate the twist-2 and twist-3 contribution in the FCSA.
To find the target mass dependence of the  spin dependent 
structure functions we have to construct the traceless nucleon matrix 
elements of above operators using  two vectors, the nucleon momentum $P$ and 
the 
spin vector $S$. The $\Theta$-operators are traceless in the massless quark
 limit.
As an example, we present here the expression
 for such a tensor for the twist-2 part of the quark operator
\begin{eqnarray}
&&\hspace{-0.7cm}\langle PS|\Theta_S^{\pm\mu_1\{\mu_2\cdots\mu_n\}}|PS\rangle = 
~~~~~~~~~~~~~~~~~~~~~~~~~~~~~~~~~~~~~~~~~~~~~~~~~~~~~\nonumber \\
&&\hspace{-19pt}a_{n-1}^{\pm}\frac{1}{n}\sum_{j=0}
{\frac{(-1)^j}{2^j}\frac{(n-j)!}{n!}\underbrace{g\cdots g}_{j}~
\underbrace{[SP\cdots P]_S}_{n - 2j}
M^{2j}}.~~~~~~~~~~~~~~~~~~~~~~~~~~~~~~~~~~~~~~~~\nonumber 
\end{eqnarray}
The  sum contains $j$ times the metric tensors $g_{\mu_i\mu_k}$.
The remaining $n-2j$  indices are symmetrized in the product 
$[SP\ldots P]_S$;
$a_{n}^{\pm}$ denotes the twist-2 operator
 reduced matrix elements. The same 
expressions can be written for the twist-3 part of the 
$\Theta$-operator using the
reduced matrix elements $d_{n}^{\pm}$. These are non-perturbative quantities 
and are independent of the nucleon mass. All information about the target
mass corrections is contained in the  tensor structures of the nucleon matrix 
elements.
Taking the nucleon matrix elements of the $T$-product of two currents one 
obtains the FCSA.  The dispersion relations result into
expressions of the moments of structure functions which are expressed 
through series of the ratio $(M^2/Q^2)^k$ (see details in \cite{BT}).
In many practical applications, as the analysis of experimental data, the
expressions for the moments of the deep inelastic SF  are less 
suited than the corresponding $x$-space expression.
To obtain these we apply  we apply  the inverse Mellin transform to the 
moments of the corresponding SF. For the twist-2 parts of the 
polarized SF we have
\begin{eqnarray}
g_1^{\pm~tw2}(x)& = &
x\frac{d}{ dx} x\frac{ d}{d x}
\left[\frac{x}{y} \frac{G^{\pm}_1(\xi)}{\xi}\right],
\label{g1G}\\
g_2^{\pm~tw2}(x)& = &-
x\frac{d^2}{ dx^2}x
\left[\frac{x}{y} \frac{G^{\pm}_1(\xi)}{\xi}\right],
\label{g2G}\\
g_3^{\pm~tw2}(x) & = & 
2 x^2\frac{d^2}{ dx^2}
\left[\frac{x^2}{y} \frac{G^{\pm}_2(\xi)}{\xi^2}\right],
\label{g3G}\\
g_4^{\pm~tw2}(x) & = & -
x^2\frac{d}{ dx}x\frac{d^2}{dx^2}
\left[\frac{x^2}{y} \frac{G^{\pm}_2(\xi)}{\xi^2}\right],
\label{g4G}\\
g_5^{\pm~tw2}(x) & = & -
x\frac{d}{ dx}
\left[\frac{x}{y} \frac{G^{\pm}_3(\xi)}{\xi}\right]
\nonumber \\
&+&\frac{M^2}{Q^2}x^2\frac{d^2}{ dx^2}
\left[\frac{x^2}{y} \frac{G^{\pm}_2(\xi)}{\xi}\right],
\label{g5G}
\end{eqnarray}
where $y = \sqrt{1+4M^2 x^2/Q^2}$ and $\xi$ is the Nachtmann variable,
$\xi=2x/\left[1+(1+4M^2 x^2/Q^2)^{1/2}\right]$
\cite{Nachtmann}.
The operator expectation values  $a^{\pm}_n$ are the moments of  
distribution functions $F^{\pm q}(x)$, which are related to the
polarized parton densities in the massless limit, $\Delta q(x) \pm
\Delta \overline{q}(x)$,
\begin{eqnarray}
a_n^{\pm,q} = \int_{0}^{1}{dy y^n  F^{\pm q}(y)}~.
\label{an}
\end{eqnarray}
The functions $G^{\pm}_i$ are related to the
distribution function $F^{\pm q}(y)$ by
\begin{eqnarray}
&&\hspace{-16pt}G^\pm_1(y) = \nonumber\\
&&\sum_q\frac{(g_V^q)^2+(g_A^q)^2}{4}
\int_{y}^{1}{\frac{dy_1}{y_1}\int_{y_1}^{1}{\frac{dy_2}{y_2}
F^{\pm q}(y_2)}}~,
\nonumber\\
&&\hspace{-16pt}G^{\pm}_2(y)  = \nonumber\\
&&\sum_q {~g_V^qg_A^q 
\int_{y}^{1}{dy_1\int_{y_1}^{1}{\frac{dy_2}{y_2}{\int_{y_2}^{1}
{\frac{dy_3}{y_3}F^{\pm q}(y_3)}}}}}~,\nonumber\\
&&\hspace{-16pt}G^{\pm}_3(y)  =  \sum_q {\frac{1}{2}~g_V^qg_A^q
\int_{y}^{1}{\frac{dy_1}{y_1}F^{\pm q}(y_1)}}~.\nonumber
\end{eqnarray}
In the same way the moments of the corresponding twist-3 contributions 
can be inverted. One obtains the following $x$-space expressions:
\begin{eqnarray}
 g^{\pm~tw3}_{1}(x,Q^2) &\hspace{-8pt}=& \frac{4 M^2}{Q^2} x^2\frac{d^2}{dx^2}
\Biggl[\frac{x^2}{y} H^{\pm}_1(\xi)\Biggr],
\label{g1t3dif}\\
 g^{\pm~tw3}_{2}(x,Q^2) &\hspace{-8pt}=&  x\frac{d^2}{dx^2}
\Biggl[\frac{x}{y} H^{\pm}_1(\xi)\Biggr],
\label{g2t3dif}\\
g^{\pm~tw3}_{3}(x,Q^2) &\hspace{-8pt}=& 
-\left(x^2\frac{d^3}{dx^3}+4\frac{M^2x^2}{Q^2}x\frac{d^2}{dx^2}x\right)
\nonumber \\
&&~~~~\times\left[\frac{x^2}{y}
\frac{H_2^{\pm}(\xi)}{\xi}\right],
\label{g3t3dif}\\
 g^{\pm~tw3}_{4}(x,Q^2) &\hspace{-8pt}=& -4 \frac{M^2}{Q^2}
x^3\frac{d^3}{dx^3}
\Biggl[\frac{x^3}{y} \frac{H^{\pm}_2(\xi)}{\xi}
\Biggr],
\label{g4t3dif}\\
 g^{\pm~tw3}_{5}(x,Q^2) &\hspace{-8pt}=& -2 \frac{M^2}{Q^2}
x^2\frac{d^2}{dx^2}
\Biggl[\frac{x^2}{y} \frac{H^{\pm}_2(\xi)}{\xi}
\Biggr].
\label{g5t3dif}
\end{eqnarray} 
The matrix elements of the twist--3 operators are the moments of the
distribution function $D^{\pm q}(x)$ in the massless limit, 
\begin{eqnarray}
d^{\pm q}_n=\int_{0}^{1}dx x^n D^{\pm q}(x)~,
\end{eqnarray}
which has, however, no partonic interpretation.

\vspace{-1cm}
\begin{figure}[htb]
\begin{center}
\includegraphics[width = 7 cm]{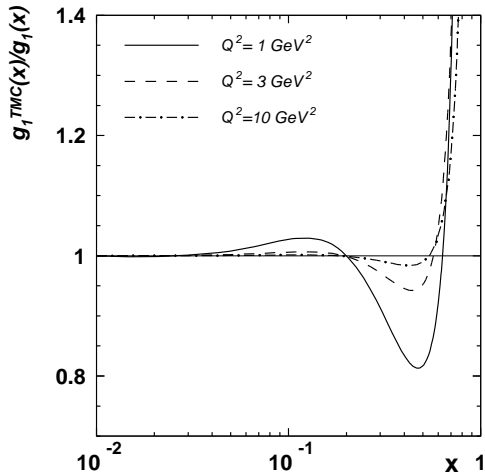}
\end{center}
\vspace{-1.5cm}
\caption{The ratio $g_1^{TMC}(x)/g_1(x)$ versus $x$.}
\end{figure}

\vspace{-0.5cm}
\noindent
The functions $H^{\pm}_{1,2}(x)$ can be expressed through   
integrals over the distribution function $D^{\pm q}(x)$ as in the twist-2 case
\cite{BT}.
Performing the derivatives in Eqs.(\ref{g1G})-(\ref{g3G}) and 
(\ref{g1t3dif})-(\ref{g5t3dif}) we obtain the final expressions for the
structure functions in $x$-space. The twist-2 and twist-3 parts are expressed
through the  $F^{\pm q}$ and $D^{\pm q}$, respectively. 
The corresponding expressions are given in Ref. [7]. 
Here we present numerical results  on the 
size of target mass effects for the twist-2 contributions to 
$g_1$ and $g_2$.
 We calculate  $g_1(x)$ 
 using the LO parametrization for polarized SF \cite{GS}.
To estimate the size of nucleon mass corrections for the twist-2 part of 
$g_1(x)$ we use the same expression for $g_1(x)$ as $F^q(x)$. 
The ratio of $g_1^{TMC}(x)/g_1(x)$ is shown in Fig. 1 for different values
 of $Q^2$.  
\begin{figure}[hbt]
\vspace{-1cm}
\begin{center}
\includegraphics[width = 7 cm]{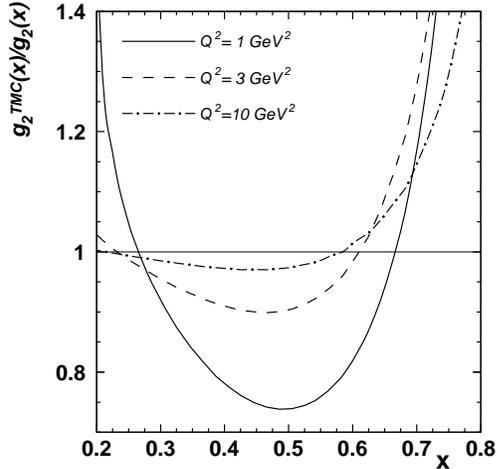}
\end{center}
\vspace{-32pt}
\caption{The ratio $g_2^{TMC}(x)/g_2(x)$ versus $x$.}
\end{figure}
The same calculation is done for the structure function $g_2(x)$,  Fig. 2.
Here $g_2(x)$ was calculated from $g_1(x)$
 using the Wandzura-Wilczek relation \cite{WW}.

\section{RELATIONS BETWEEN THE \\  STRUCTURE FUNCTIONS}

As was shown in a previous analysis~\cite{BK} the twist-2 contributions
to the polarized structure functions are connected by three (integral) 
relations, the Dicus--relation~\cite{DIC}, the Wandzura--Wilczek (WW) 
relation~\cite{WW} and a new relation by Bl\"umlein and Kochelev \cite{BK}.
It can easily be seen from Eqs. (2)-(6) that 
the nucleon mass correction does not affect the WW relation as well as 
the relation of Ref. \cite{BK}. 
Moreover, the WW relation is not violated even by quark mass corrections 
\cite{BT}. The  Dicus relation is violated in the presence
of nucleon mass corrections 
similarly to the Callan--Gross relation~\cite{CG} in the unpolarized
case.

In the presence of target mass corrections all structure functions 
$\left. g_i\right|_{i=1}^5$ contain twist--3 contributions. On the 
contrary, in the massless limit this is the case for the structure
functions $g_2$ and $g_3$ only~\cite{BK}.
 The mass dependence of the scattering
cross sections for longitudinal nucleon polarization, 
however, reveals that both the structure functions $g_2$ and $g_3$ do 
only contribute at $O(M^2/Q^2)$. Therefore, a complete account for 
twist--3 contributions requires to consider also the nucleon mass 
corrections for the other polarized structure functions.
 From Eqs.~(\ref{g1t3dif}--\ref{g5t3dif}) one derives
the following {\it new} relations between the twist--3 parts of 
the different spin--dependent structure functions~:

\vspace{-0.5cm}
\begin{eqnarray}
&&\hspace{-16pt} g^i_{1}(x,Q^2)  = \nonumber \\
&&\hspace{-16pt}~~ \frac{4 M^2 x^2}{Q^2}
\left[ g^i_{2}(x,Q^2)
-2\int_{x}^{1}{\frac{dy}{y} g^i_{2}(y,Q^2)}\right],
\label{t3g1g2}\\
&&\hspace{-16pt}\frac{4 M^2 x^2}{Q^2} g^i_{3}(x,Q^2) =  
g^i_{4}(x,Q^2)\left(1+\frac{4 M^2 x^2}{Q^2}\right) \nonumber\\
&&~~~~~~~~~~~~~~~~~~~~~~~ +3\int_{x}^{1}{\frac{dy}{y} g^i_{4}(y,Q^2)},
\label{t3g3g4}\\
&&\hspace{-16pt}2 x g^i_{5}(x,Q^2) 
=  -\int_{x}^{1}{\frac{dy}{y} g^i_{4}(y,Q^2)}.
\label{t3g4g5}
\end{eqnarray}

\vspace{-0.2cm}
\noindent
Here, Eq.~(\ref{t3g1g2}) at the one side, and 
Eqs.~(\ref{t3g3g4},\ref{t3g4g5}) on the other side correspond to
different flavor combinations among the twist--3 contributions. This is
similar to the case of the twist--2 terms, where the former case
corresponds to the combination $\Delta q + \Delta\overline{q}$, 
and the latter to $\Delta q - \Delta\overline{q}$.

 Eqs.~(\ref{t3g1g2}--\ref{t3g4g5}) show that the twist--3
contributions to $g_1, g_4$ and $g_5$ vanish in the limit $M \rightarrow
0$. On the other hand, if one keeps terms of 
$O\left[(M^2/Q^2) \cdot g_{2(3)}\right]$ 
and twist--3 in the scattering cross sections, one has to account also for
the twist--3 terms in $g_1, g_4$ and $g_5$.

\section{CONCLUSION}           

\vspace{2mm}
\noindent
We have calculated the target mass corrections for all polarized structure 
functions for both neutral and charged current deep inelastic scattering.
The results were obtained by using the
local light cone expansion of the FCSA.
 The target mass corrections imply besides the twist--2 terms 
twist--3 
contributions for all polarized structure functions. 
 The corrections were both represented in terms of the integer moments which 
result from the light cone expansion and their analytic continuation and 
Mellin inversion to $x$-space. 
The size of nucleon mass corrections is expected to be  $20\%-40\%$ for
 $Q^2\simeq 1$ GeV$^2$ at moderate values of $x$ 
for the twist-2 contributions of
$g_1(x)$ and $g_2(x)$.

We investigated the effect of the target mass corrections on the sum rules
connecting the polarized structure functions in lowest order in the
coupling constant. For the twist--2 contributions both the 
Wandzura--Wilczek relation~\cite{WW} and the relation derived in 
Ref.~\cite{BK} are preserved, whereas the Dicus relation~\cite{DIC}
 receives a correction. It was also shown that the Wandzura--Wilczek
relation is preserved in the presence of quark--mass corrections. 

Three new integral relations were derived for the twist--3 contributions 
of the polarized structure functions. They hold without further assumptions
on the flavor combinations of the related structure functions.

{\bf Acknowledgment.}
We would like to thank W.L. van Neerven for discussions. The work was 
supported in part by EU contract FMRX-CT98-0194(DG 12 - MIHT) and the 
Alexander von-Humboldt Foundation.



\begin{thebibliography}{999}
%
\bibitem{GP}
H.~Georgi and H.D.~Politzer, Phys.~Rev. {\bf D14} (1976) 1829.
%
\bibitem{PR}
A.~Piccione and G.~Ridolfi, Nucl.~Phys. {\bf B513} (1998) 301.
%
\bibitem{WA}
S.~Wandzura, Nucl.~Phys. {\bf B122} (1977) 412.
%
\bibitem{MU}
S. Matsuda and T. Uematsu, Nucl. Phys. {\bf B168} (1980) 181.
%
\bibitem{BK}
J.~Bl\"umlein and N.~Kochelev, Phys. Lett. {\bf B381} (1996) 296;
Nucl.~Phys. {\bf B498} (1997) 285.
%
\bibitem{Nachtmann}
O.~Nachtmann, Nucl.~Phys. {\bf B63} (1973) 237.
%
\bibitem{BT}
J.~Bl\"umlein and A.~Tkabladze, DESY 98-181, 
Nucl.~Phys. {\bf B} in print, hep-ph/9812478.
%
\bibitem{WW}
S.~Wandzura and F.~Wilczek, Phys. Lett. {\bf B72} (1977) 195.
%
\bibitem{DIC}
D.A.~Dicus, Phys. Rev. {\bf D5} (1972) 1367.
%
\bibitem{GS}
T.K.~Gehrmann and W.J.~Stirling, Phys.~Rev. {\bf D53} (1996) 6100.
%
\bibitem{CG}
C.G.~Callan and D.J.~Gross, Phys. Rev. Lett. {\bf 22} (1969) 156.
\end{thebibliography}
\end{document}